%% file: main.tex
\documentclass[a4paper]{article}
\usepackage[backend=biber]{biblatex}
\usepackage[hidelinks]{hyperref}
\usepackage{booktabs}
\usepackage{pdflscape}
\usepackage{afterpage}
\usepackage{setspace}
\usepackage{xcolor}
\usepackage{enumitem}
\usepackage{amsthm}

\title{Sharing and Caring\\[6pt]\normalsize Creating a Culture of Constructive Criticism in Computational Legal Studies}
\author{Corinna Coupette and Dirk Hartung}
\date{}

\newtheorem{principle}{Principle}

\bibliography{bibliography}

\begin{document}
	\onehalfspacing
	\maketitle
	\begin{abstract}
		\input{text/abstract}
	\end{abstract}
	
	\input{text/introduction}

	\input{text/daring}

	\input{text/sharing}

	\input{text/caring}

	\input{text/outlook}	
	
	\printbibliography
	
\end{document}

%% file: text/abstract.tex

\noindent 
We introduce seven foundational principles for creating a culture of constructive criticism in computational legal studies.
Beginning by challenging the current perception of \emph{papers} as the primary scholarly output, 
we call for a more comprehensive interpretation of \emph{publications}.
We then suggest to make these publications \emph{computationally reproducible}, 
releasing all of the data and all of the code all of the time, on time, and in the most functioning form possible.
Subsequently, we invite \emph{constructive criticism} in all phases of the publication life cycle.
We posit that our proposals will help form our field, 
and float the idea of marking this maturity by the creation of a modern flagship publication outlet for computational legal studies.

%% file: text/introduction.tex
\section{Introduction}

Code and data unavailable, available upon ``reasonable request'', or from dead links only. 
Little, if any, documentation of underlying assumptions or judgment calls. 
Lack of sensitivity analyses, robustness checks, or ablation studies. 
Limited peer review, or peers impressed by figures showing results produced by algorithms they do not fully understand, on data whose provenance is unclear. 
Referenced sources behind paywalls---%
or not indexed by common search engines at all. 
The list of deficiencies affecting published papers in computational legal studies goes on. 
How come? 

The answer is simple, yet unsettling: 
\emph{Computational legal studies} (CLS), 
broadly defined as \emph{the study of law using computational methods}, 
is hard. 
Things can go wrong easily. 
Misspecified models, dirty data, buggy code. 
No individual researcher is perfect, but as a community, we can strive to identify our mistakes, correct them, and learn from them for the future. 
We can get better, individually and collectively, and we can make progress. 
This, however, requires scientific hygiene routines that have yet to be established. 
As our research develops at the intersection of law and computer science, 
and articles using computational methods make their way into mainstream legal research outlets, 
we can no longer ignore the striking mismatch between the publication procedures familiar from doctrinal scholarship and empirical legal studies on the one hand, 
and the requirements of robust, reproducible computational legal research on the other.

In this essay, 
we argue that for computational legal studies to advance as a community, 
the field needs a publication culture designed to meet its unique challenges. 
We find the building blocks of such a culture in our parent disciplines. 
From computer science, we can adopt the increasingly widespread requirements of data availability, code availability, honest assessments of the methodological and interpretive limitations of our research, and transparent, constructive criticism of our own work and the work of others. 
Legal publication culture offers other advantages: 
Less driven by conference deadlines and less overwhelmed by mass peer review, 
legal scholars can make time to focus on big ideas, rather than merely pushing for incremental improvements. 
Hence, combining the best of both our worlds can help us keep our studies both scientifically rigorous and comprehensible for a heterogeneous audience comprising both legal scholars and computer scientists.

As this essay is about the culture of our field, it is first and foremost an invitation for discussion.
We care about creating a constructive and critical community culture, 
and we share our ideas on how to get there,
but we do not pretend to know better than our fellow researchers.
Every scientific debate requires a starting point---%
a set of ideas to be criticized, improved, and ultimately either adopted or rejected.
We hope that the remainder of this publication will spark productive controversies.
A final disclaimer: 
All of our suggestions are born from experience, 
and we ourselves have sometimes fallen short on some of our suggestions. 
We are not above making mistakes, 
and we, too, have known better just after a work was published.
But we constantly try to improve, 
which is what motivated us to write this piece. 
You are more than welcome to join us for the ride.

%% file: text/daring.tex
\section{Daring: Challenging Current Conceptions}
\label{sec:daring}

In both computer science and legal scholarship, papers and publications are often used as synonyms. 
Papers may include preprints, 
while publications are papers that were submitted and accepted at a venue where they could also have been rejected (e.g., a conference or a journal). 
In common usage, the implied content of papers and publications, however, is the same:
Legal scholars expect mostly natural language text, ideally including many footnotes,
and computer scientists expect a (domain-dependent) mixture of natural language text, math, algorithms formulated in pseudocode, tables, figures, and references.

In their composition, computational legal studies come closer to \emph{computer} science publications (e.g., \emph{computable law})  or \emph{computational} science publications (e.g., \emph{legal data science})
than to doctrinal legal publications. 
CLS are more than cogent arguments crafted in natural language, and hence, 
they have more potential points of failure than doctrinal legal publications: 
\emph{We can be wrong in many ways}. 
Fortunately, though, especially where our studies involve math, algorithms, code, or data,  
\emph{we can be proven wrong}, 
which allows us to discover mistakes, correct them, and make progress.
Similarly, 
where our work develops methods whose performance on certain tasks can be assessed systematically, 
\emph{we can be suboptimal in many ways}, 
but the upside of this is that \emph{we can measurably improve}. 

As errors and imperfections are inevitable, 
we must strive to sustain a spiral of community self-correction and advancement to safeguard scientific progress.  
To achieve this, we need to embrace two concepts:
\emph{computational reproducibility} and \emph{constructive criticism}.
\emph{Computational reproducibility} means reproducing reported results using the same input data, computational steps, methods, and code \cite{national2019reproducibility}, 
and it implies that the necessary materials are available. 
As such, it is a prerequisite for \emph{constructive criticism}, 
i.e., the critical checking of studies with a view to improving them.
This leaves us with two questions:
\begin{enumerate}[itemsep=0em]
	\item How do we ensure that our results are \emph{computationally reproducible}?
	\\($\rightarrow$~\ref{sec:sharing}: \emph{Sharing})
	\item How do we ensure that our results are \emph{constructively criticized}?
	\\($\rightarrow$~\ref{sec:caring}: \emph{Caring})
\end{enumerate}
While we might not address these questions for all of science, 
we \emph{can} develop best practices for the CLS community. 
The first step in this endeavor is to \emph{reset our terminology}, 
using paper and publication for different things.
A \emph{paper} is exactly what is currently synonymously understood by paper \emph{and} publication.
A \emph{publication} includes a paper, 
but beyond that, it \emph{must} comprise all materials required to computationally reproduce the results reported in the paper 
(first and foremost: \emph{data} and \emph{code}), 
and it \emph{may} further contain presentation materials (e.g., slides, videos, and posters) 
and additional text elaborating on points from the paper (e.g., the classic \emph{supplementary information}). 
In brief:
\begin{center}
	Paper $\neq$ Publication. Paper $\in$ Publication.
\end{center}
We should create and criticize publications, not papers.

%% file: text/sharing.tex
\section{Sharing: Really Reproducible Research}
\label{sec:sharing}

The first prerequisite of sustainable scientific progress in CLS is \emph{computational reproducibility}.
Targeting this goal forces us to face an open problem familiar from computer science and computational science: 
How do we deal with the data underlying the figures and tables, 
workflows, pseudocode implementations, 
and the code analyzing, tabulating, or visualizing the data---%
without which it is impossible to computationally reproduce reported results or discover mistakes?
As summarized succinctly by \citeauthor{buckheit1995wavelab} \cite{buckheit1995wavelab} paraphrasing Stanford seismologist and \emph{really reproducible research} champion \emph{Jon Cl\ae rbout} in the context of \emph{computational} science (emphasis in the original):
\begin{quote}
	``\emph{[a]n article about computational science in a scientific publication is \textbf{not} the scholarship itself, it is merely \textbf{advertising} of the scholarship. 
		The actual scholarship is the complete software development environment and the complete set of instructions which generated the figures.}'' 
\end{quote}
So, how do we ensure that we publish scholarship, not just advertising?
Acknowledging that more detailed guidelines exist in the literature on computational reproducibility \cite{krafczyk2021learning},
we propose to begin with the following simple principles:
\begin{principle}
	Release your data.
\end{principle}
\begin{principle}
	Release your code.
\end{principle}

To put these principles into practice, 
we need to answer three questions: 
(1)~\emph{What} data and code should be released,
(2)~\emph{how} should they be released, 
and (3)~\emph{when} should the release happen?
In the following, we address each of these questions, 
also responding to some \emph{$\dots$but what if} (counter-)questions that we have encountered in our own research and reviewing practice.

\subsection{\emph{What} data and code should be released?}

The initial answer to this question is simple: 
Ideally, \emph{all} data and code should be released. 
The contents of this requirement differ by resource type.

\emph{Data} includes the \emph{raw data} (as obtained from its original source, e.g., a website or a database), 
the \emph{preprocessed data} underlying our analysis, 
the \emph{result data} obtained from running algorithms,
and any \emph{postprocessed data} underpinning our figures and tables.
A classic \emph{$\dots$but what if} question arising in this context concerns data to which legal restrictions apply 
(e.g., because the data are sensitive or were obtained from a commercial data provider under a non-disclosure agreement). 
If the results of a study hinge on such data, 
we should obviously abide by the law, 
but we should opt for a narrow construction thereof in the short term
(e.g., where possible, construing it such that citation networks derived from legal documents can still be shared),
explore options to share data under access restrictions in the medium term (e.g., data might be unlocked for scientific purposes such as peer review only),
and advocate for an improvement of the legal situation surrounding our data in the long term.
Another (perhaps subconscious) worry in the context of data release relates to poor data quality.
However, if there are known data quality limitations or concerns,
these should already be documented in the first paper using the data, 
or in the materials supplementing that paper. 
The community is well aware that getting the data exactly right is almost impossible, 
and will not condemn those who, for example, fail to extract a few oddly formatted citations. 
Any remaining quality-related hesitancy by the authors to release the data underlying a computational legal study should be treated as a warning sign 
and cause them to revisit the data, rather than proceed to publish their paper.

\emph{Code} includes the libraries, scripts, and notebooks used in all stages of the data life cycle \cite{berman2018realizing} (for legal data science),
i.e., to acquire, clean, use, communicate, and archive the data;
programming language artifacts (for computable law); 
and a specification of the computational environment (including dependencies on external software) facilitating its recreation. 
Here, a common \emph{$\dots$but what if} question concerns code quality, 
or rather, a perceived lack thereof. 
If we feel that our code is not in good shape, we should keep in mind that public replication code in bad shape is still much, much better than no public replication code at all. 
We all know what can be reasonably expected from research code 
(with humor, see the Community Research and Academic Programming License, aptly abbreviated as \hyperref{https://matt.might.net/articles/crapl/}{}{}{CRAPL}), 
and hence, will not judge anyone for their lack of unit tests or documentation.
As the community will still appreciate code in good shape, 
in this situation, 
we should promise ourselves to pursue better practices in the future 
(for basic guidance, see the ten principles laid out in \cite{hunter2021ten}).
A particularly concerning variant of the \emph{bad code shape} argument is that \emph{code may not exist}, 
e.g., because the authors conducted their analyses using a button-click spreadsheet program. 
This makes it practically impossible to ensure computational reproducibility, 
and it keeps leading to high-profile failures (for an example, search ``Rogoff Reinhart Excel'').  

Finally, one \emph{$\dots$but what if} question arising for both data and code concerns perceived \emph{triviality}: 
\emph{$\dots$but what if} we think that the data or code to be released are readily available from elsewhere, 
or easy for others to come up with themselves?
To the extent (and \emph{only} to the extent) that they have already been released under persistent \emph{Digital Object Identifiers} (DOIs), 
we should cite their DOIs instead of redistributing the data or code ourselves, thereby giving credit to their original creators. 
In all other cases, it should not be a problem to release our trivial materials anyways, right?
The purpose of data and code releases is to enable others to assess 
if (and how) \emph{what we actually did} corresponds to \emph{what we said we did} in our papers, 
regardless of how complicated that was,
and to complete this check with reasonable effort. 
Furthermore, databases change in composition, public data sources are moved or removed,
and reading code is typically much faster than writing it.

\subsection{\emph{How} should the data and code be released?}

To start with the obvious:
Data and code should neither be shared \emph{upon request} only, 
nor by making them available from a personal or institutional website.
Even if we hope otherwise, authors and websites may move, die, or simply become unresponsive \cite{tedersoo2021data}. 
We should go beyond current practices in computer science,
where disclosing the data and code informing a paper is becoming increasingly common, 
but practices vary across and within domains, institutions, and research groups, 
and the quality of releases is often low. 

Ideally, data and code should be released such that they are findable, accessible, interoperable, and reusable---%
i.e., FAIR in the sense of the \hyperref{https://www.go-fair.org/fair-principles/}{}{}{FAIR Principles} \cite{wilkinson2016fair,lamprecht2020towards}.
To ensure computational reproducibility, \emph{findability} and \emph{accessibility} are most important.
While it is easy to get lost in the detailed debates around FAIR sharing (see, e.g., the \hyperref{https://www.fairsfair.eu/}{}{}{FAIRsFAIR Project}),
a convenient way to achieve minimum compliance with the FAIR principles 
is to deposit data and code with an archiving service that adheres to these standards, 
such as \hyperref{https://about.zenodo.org/principles/}{}{}{Zenodo}.
Deposits created by such a service are immutable after publication, 
come with required metadata (which is mutable after publication), 
and are assigned a persistent DOI, 
which can be referenced in the papers that are based on these deposits, 
and which can be resolved reliably via the DOI system.
Changes to the files included in a release can be published as a new version of the original deposit and connected to that deposit via DOI versioning.

Moving beyond minimum compliance with the FAIR principles, 
we can ask how to structure our releases to optimize for \emph{reusability}, 
which can help hasten community progress.
Here, we must acknowledge that while a publication should be immutable,
datasets evolve, 
and most code must be maintained to remain runnable.
Therefore, we should release the reusable parts of our data and our code separately, 
such that we may create new versions of them independently. 
We could then craft a \emph{wrapper release} for our publication,
which would link to the versions used to produce the results reported in the associated paper, 
and which would contain all intermediate data and code produced specifically to write that paper, 
plus (potentially) the source of the paper. 

This practice allows us to integrate archiving services directly with version control services 
while following best practices for code maintenance using a popular developer platform
(e.g., Zenodo offers \hyperref{https://docs.github.com/en/repositories/archiving-a-github-repository/referencing-and-citing-content}{}{}{GitHub integration} in the sense that a GitHub release can trigger a corresponding Zenodo release). 
Similarly, our dataset releases can follow best practices for dataset documentation, 
such as \emph{datasheets for datasets} or \emph{dataset nutrition labels} \cite{gebru2021datasheets,holland2020dataset,chmielinski2022dataset}.
The primary responsibility of the publication-specific wrapper release, then, 
is to tie all resources together, such that the results from its associated paper can \emph{really} be reproduced. 
Here, releasing data and code as described above may not be enough to guarantee computational reproducibility. 
Rather, the entire \emph{workflow} should be documented and included in the wrapper release, 
ideally in the form of a script that, upon execution, 
produces the paper from the other contents of the release. 
This is a variation of the idea of \emph{research compendiums}, 
i.e., executable research objects that combine paper, data, code, environment, and narrative \cite{gentleman2007statistical,brinckman2019computing}, 
which is gaining traction in computational science.
The ideas underlying research compendia can be traced back to \emph{Donald Knuth}'s ideas on literate programming \cite{knuth1984literate}, 
but promising systems enabling their easy creation are still prototypes 
(e.g., \hyperref{https://wholetale.org/}{}{}{Whole Tale}, an NSF-funded project initiated in $2016$, states ``expected completion in February,  $2022$''),
and it will take years, if not decades, until they are widely adopted. 
Hence, in the meantime, we complement our first two principles with a third principle, 
whose implementation can be inspired by guidelines from other disciplines \cite{stoudt2021principles}: 
\begin{principle}
	Release your workflows.
\end{principle}

\subsection{\emph{When} should the data and code be released?}

As a rule of thumb, all data and code used in a publication should be released with the first non-preprint paper using them.
This should be self-evident, 
given that it is necessary to ensure that published results are computationally reproducible.
We highlight it because we have seen others raise a particularly unnerving \emph{$\dots$but what if} question concerning release timing:
\emph{$\dots$but what if} putting together the data or code was a lot of work, or there is other work in the pipeline that uses the data or code?

Acknowledging that this might be controversial, 
we hold that there is no legitimate argument to withhold the data or code underlying a scientific publication in our field 
on the grounds that putting them together was effortful, 
or that we plan to continue working with them in the future. 
Creating, monopolizing, and exploiting resources prioritizes the scientist, not the science. 
People can do it, but we should not reward them by publishing their papers (which, following the terminology from Section~\ref{sec:daring}, are really only papers, not publications).

If putting together our data or code was laborious, 
or we want to use them in the future, 
both of which are rather rules than exceptions,
then it is even more crucial that we allow others to validate and leverage our resources.
Research is a collaborative endeavor, 
and withholding data or code in the hope of exhausting them without ``competition''{} 
is contrary to the ethos of science with its norms of universalism, \emph{communalism}, \emph{disinterestedness}, and organized skepticism \cite{merton1979normative}.

%% file: text/caring.tex
\section{Caring: Constructive Community Criticism}
\label{sec:caring} 

The second prerequisite of sustainable scientific progress in CLS is \emph{constructive criticism}.
This criticism is desirable in all phases of the publication process: 
pre-publication, in-publication, and post-publication. 

\subsection{Pre-Publication Criticism}
\label{subsec:prepublication}

Despite its imperfections in practice \cite{gerwing2020quantifying,gerwing2021reevaluation,tennant2020limitations}, 
a culture of peer review is the cornerstone of constructive criticism in the pre-publication phase.
This is the point in the publication cycle at which we can shape our field most effectively, 
since authors eager to publish their work are most motivated to invest additional resources to implement improvements.
All our research should go through peer review 
because it offers the opportunity to receive constructive criticism for our work and ameliorate our scholarship.
For this reason alone, it is easy to cheer for peer review.
But not peer review of any kind. 
Three important questions remain:
(1)~\emph{What} should be peer reviewed, 
(2)~\emph{who} should conduct the peer review, and 
(3)~\emph{how} should the peer review be performed?

\subsubsection{What should be peer reviewed?}

Traditional peer review focuses on papers, 
but CLS publications are much more than that.
Most importantly, a review of a CLS \emph{paper} alone does not suffice 
to ensure computational reproducibility and assess the merits of the methodology employed.
The CLS review process, therefore, should encompass all required elements of a \emph{publication} as defined in Section~\ref{sec:daring}.

Not only are data and code necessary to spot bugs, 
which---among other things---is exactly what peer review is for.
By not requiring data and code in the review process, 
we actually devalue the labor that goes into these parts of a publication.
In CLS, as in many areas of computer science and computational science, 
researchers regularly spend most of their time wrangling data or writing code.
Since legal data are seldom easily accessible, 
the resources invested to curate datasets should be appreciated and highlighted.

Especially when reviewing computational legal studies 
for outlets unfamiliar with the best practices in computer or computational science,
it can be hard for peers to obtain the data and code underlying a study they ought to review. 
Here, editorial policies embracing the \hyperref{https://www.cos.io/initiatives/top-guidelines}{}{}{Transparency and Openness Promotion (TOP) guidelines} \cite{nosek2015promoting} 
and, inter alia, 
requiring authors to provide data and code availability statements, 
submit their materials for review to the greatest possible extent, 
and explain why they are unable to share (parts of) their materials in the review process, 
could make our lives much easier. 
After all, if authors fail to give us access to the data and code needed to ensure computational reproducibility 
(with potential exceptions for privileged \emph{data}, in which case we can still review the \emph{code}), 
we should decline the review or desk reject the submission. 
A rigorous review process, combined with strict editorial policies, 
is the best opportunity to ensure that the CLS community creates publications, not just papers. 

\subsubsection{\emph{Who} should conduct the peer review?}

Now that we know what to peer review: 
Who should be the reviewing peers?
Here, we need to acknowledge that our community, though growing, is currently still fairly small.
Hence, although there exist interesting alternative models, 
such as expert crowd review \cite{list2017crowd,gemmeren2021crowd}, 
for the moment, 
we should probably stick with traditional review by a small number of peers.
Furthermore, truly double-blind peer review seems hardly possible, 
especially in light of prevalent citation practices.
Instead, reviewers should simply indicate whether they recognize the authors of a publication and provide their level of certainty, 
thus allowing editors to make an educated decision.

We trust that when embedded in the right culture,
partial or full identification will not hinder a productive collaboration in the review process.
To counter the risk of \emph{negative} social ties getting in the way of objective assessments, 
authors could be given the option to exclude certain reviewers from the start.
In (somewhat) single-blind peer review, however,
\emph{positive} social ties might become problematic.
Here, the easiest solution is to exclude reviewers who collaborate with the authors, 
work at the same institution, or are otherwise academically related (e.g., via PhD supervision),
in a certain time window before or around the review.
As our collaboration network becomes increasingly dense, however,
this might be easier said than done.
An alternative approach would be to put the responsibility to judge their capacity to objectively evaluate a publication with individual reviewers.
While reviewers might occasionally have incentives to brush aside personal bias,
communicating the review process transparently (as detailed below) could alleviate some of these concerns.
In our view, it is much more important to have qualified reviewers who properly understand the authors' problems and methods 
than reviewers who have absolutely no academic ties to the authors but also little expertise in the relevant domain because they are working in a tangentially related area.
In our growing field, 
the problem will likely solve itself over time. 
Until then, though, 
strictly prohibiting reviewers with prior collaborations 
risks substantially increasing the number of reviews to be conducted by less connected (and perhaps also less experienced) scientists, 
which would be detrimental to the development of our field.

Taking into account both the extensive scope of reviews sketched above and the growth of our field,
there is a real risk of overextending reviewers \cite{aczel2021billion}.
This is already a common phenomenon at large computer science conferences, 
and the consequence is clear: 
As the number of reviews increases, 
the thoroughness and the overall quality of reviews decrease.
Extending review deadlines or distributing reviews over time via rolling review (e.g., as piloted by \hyperref{https://aclrollingreview.org/}{}{}{ACL Rolling Review}) are hardly solutions,
as there are \emph{always} myriad projects competing for scientists' time and attention.
To ensure high-quality review, 
we should rather find a way to motivate reviewers beyond their intrinsic interest in advancing science.

The strongest extrinsic motivation would come from recognizing reviews as serious scholarly contributions 
and acknowledging them accordingly in applications for academic positions and performance reviews.
This requires that reviews be made evaluable themselves. 
At least for reviews of publications that are eventual \emph{accepts},
the solution is relatively straightforward:
All communications between authors and (identified) reviewers are published alongside the work in question,
adopting an \emph{open review} approach in the broadest possible sense \cite{ross2017open}. 
The resulting transparency will motivate both authors and reviewers to collaborate in a productive manner, improving the scientific content of the reviewed publication, 
and institutions will have an easy way to get an impression of a candidate's reviewing activity.

The procedure for eventual \emph{rejections}, however, is more delicate.
In an ideal world, 
the culture of our field would enable an equally open and transparent communication. 
In the current state, however, resulting social frictions risk impacting the review process 
if the identities of reviewers are published after rejection. 
Furthermore, a public decision not to accept a publication might adversely affect 
its chance of being accepted even once improved or submitted to a more suitable venue.
Hence for the time being, reviewers should remain anonymous, 
and rejections should remain confidential. 
To properly reward reviewers and thereby motivate thorough reviews, 
all publication venues in our field should offer some form of elementary and advanced \emph{review recognition service}, 
e.g., 
providing written confirmation letters for performed reviews, 
and \hyperref{https://support.orcid.org/hc/en-us/articles/360006971333-Peer-Review}{}{}{adding reviews to ORCID profiles} as a trusted organization.

\subsubsection{\emph{How} should the peer review be performed?}

Knowing what should be reviewed by whom, 
what remains to be established is an understanding of the criteria for and culture of the review itself.
A good review is contextualized through two honest assessments by the reviewers themselves: 
First, we need to examine how well we understand the publication. 
We can be qualified to review a publication even if we do not understand some specific points,
but we should highlight exactly which parts of a publication we have not fully understood, 
as there is a fair chance that our lack of understanding might translate into confusion for later readers.
Second, we should be transparent about the certainty of our judgment.
Even if we understand the paper well,
we might be conflicted about its relevance, 
or we might have doubts about some of its conclusions without being able to precisely pinpoint the reasons for these doubts. 
On other issues, we might be very confident and could provide clear instructions to implement improvements. 
Both authors and editors deserve to know which of our comments falls into which category.
Since---at least in our understanding---peer review is not an adversarial process but a collaboration, there is little need to fake confidence. 
Ideally, reviewing systems would thus provide standardized scales to indicate scores for both understanding and certainty.

When it comes to the review itself,
different strategies might be in order depending on the position of the review on the \emph{publication quality spectrum}. 
Review is easy when we have publications that follow the ideals laid out above, 
i.e., they are of high quality, thoroughly researched, modestly written, and comprehensively documented. 
These publications should still be reviewed rigorously, 
but the review will be relatively straightforward, 
and it is acceptable to simply accept them.
In particular, there is no need to criticize something for the sake of critique, 
to demonstrate that we read the paper, 
or to sneak in one more reference to our own work. 
Moreover, we should let authors know when they made a great contribution. 
Research can be tough, and we can all use a pad on the back for good work every now and then.

Review is most difficult for publications that are in bad shape, yet show potential.
Unfortunately, it is these publications for which peer review is also most crucial.
As long as the core findings are correct, 
we should be generous with language, references, and structure of the paper,
and provide concise, constructive, and actionable feedback on how to improve these aspects in the revision.
Any critique should be about the publication, not the author, and come in clear, open language, combined with a friendly tone.
While we need to uphold scientific standards,
we should not needlessly scare away fellow thinkers.
The more people think about the problems in our space, 
the more likely we are to advance our understanding. 
In this spirit, let us also keep room for publications of varying sizes---%
a minor contribution is still a contribution, and it could be the entry point of a novice into the community.

Review is easy again if publications are simply bad.
Bad publications should be rejected, 
since scientific rigor has to be the core yardstick for review.
If results are unsubstantiated or methods incorrectly applied, 
this should be stated clearly, and the publication should be rejected if the authors refuse to remedy our concerns---%
just like when they are unwilling to provide the data and code needed to ensure computational reproducibility.
Here, we have to hold a firm line for both our community specifically and science more generally.
Sometimes, a promising publication in bad shape can turn into an outright bad publication 
if the authors decide not to engage in an interactive, collaborative review process.
In these cases, we should not waste time on endless discussions. 
Every communication between reviewers and authors should noticeably advance the paper.
If there is no such progress, a quick rejection is the right move.

Finally, we should strive for speed in the review process.
The vast majority of reviews can be performed in a couple of days of focused work, 
yet the review process often takes months instead of weeks. 
In a fast-moving field such as ours, this is far too long. 
If we keep publications comprehensive and accessible, 
even the extensive and thorough review suggested above should be doable over a long weekend.
If we achieve recognition of this work as serious academic output, 
it should be possible to regularly prioritize reviews for faster community progress.

All of the above is directed at scientists-as-reviewers directly. 
We believe that this is the most effective way to handle our current challenges, 
and it is the only option that is in our direct and immediate control.
Publication outlets, however, have a role to play, too.
Their reviewer selection and editorial policies affect the review process and sometimes run counter to our suggested cultural change.
Given their central position in the current publishing process and the resulting scale of changes, 
we should push them to follow the \hyperref{https://www.cos.io/initiatives/top-guidelines}{}{}{TOP guidelines} \cite{nosek2015promoting} 
and include transparency, openness, and reproducibility requirements in their guidelines for both authors and reviewers.
This can be done if we lead by example in our roles as authors, editors, and reviewers.
Therefore:
\begin{principle}
	Review rigorously.
\end{principle}

\subsection{In-Publication Criticism}

Constructive criticism should also inform the content of our publications.
Most importantly, 
we should critically assess our results in every step of our workflow.
While this applies to all of scientific research, 
our multi-step pipelines transforming raw data, often represented as natural-language text, into insights
are particularly error-prone. 
This is part of what makes CLS challenging and exciting,
but as outlined above, 
our field currently lacks best practices and robust infrastructure to effectively safeguard against unintended outcomes.
We know from other disciplines such as chemistry that even where standard tools exist \cite{willoughby2014guide},
they can still have bugs \cite{bhandari2019characterization}, 
as risks of unforeseen consequences are inherent in the complexity of computer code. 
These risk can be partially mitigated by ensuring computational reproducibility (see Section~\ref{sec:sharing}) and transparency in review (see Section~\ref{subsec:prepublication}),
but they should also motivate us to keep our claims modest and our interpretations cautious.

For example, in legal data science,
we need to ascertain that our findings really come from the data, 
i.e., that they are not artifacts of our decisions in data collection, modeling, or analysis.
To this end, we should extensively explore our model and parameter spaces before picking particular configurations. 
We should complement all analyses with sensitivity and robustness checks (potentially included in the supplementary information), 
perform ablation studies, 
document our judgment calls, 
and explain our choices.
Finally, we should highlight potential weaknesses and uncertainties afflicting our work in the text reporting its results, 
and reflect upon their implications where they are most relevant---%
and not just as an afterthought in the discussion section. 

Striking a careful tone will encourage others to critically engage with our work, 
which is one of the preconditions for progress.
As modesty signals openness to discussion, 
it will equally motivate others to reach out and share their findings on suspected shortcomings of our research.
Practicing humility will establish an atmosphere in which identifying flaws in the work of others is socially accepted and even encouraged. 
As a consequence, 
mistakes can be discovered and corrected more quickly.
Humbleness also holds another advantage specifically for CLS 
because it can reduce opposition from doctrinal legal scholars.
These researchers, who constitute an important subset of our audience, 
can easily feel threatened by our methods.
Mathematical representations, algorithms, and measurable results, 
which might make CLS appealing to us, 
tend to radiate objectivity and confidence. 
Whether intended or not, our publications are often read as proselytizing by our more traditional colleagues.
Hence, keeping our claims confined and carefully guarded against over-interpretation 
can improve acceptance among our colleagues and enable fruitful collaborations.
Thus, echoing a recent call by the editors of a prominent journal \cite{nature2020tell}:
\begin{principle}
	Tell it like it is.
\end{principle}

\subsection{Post-Publication Criticism}

Once a computational legal study is published, 
the simplest and most standard way to engage in constructive criticism is via literature reviews and carefully crafted related work sections.
If we want to achieve an actual meshing of publications, 
we should make better use of these sections and not treat them as a collection of gestures to loosely related publications.
References should be earned not by mere proximity of the topic covered 
but by publication quality, 
i.e., we should reference works that have inspired us and that we could build on.
Applying this standard will likely result in fewer references 
but improve our bibliographies nonetheless, 
as the quality and the signal of individual references increase. 
Following our holistic understanding of publications (see Section~\ref{sec:daring}),
references should also not be restricted to papers but rather extend to all components of publications, especially data and code.
Let us highlight when we were able to utilize others' materials, 
and let us be honest when we could not.

This serious engagement with prior work naturally adds to the scientific workload.
We might be able to offset some of that effort by making it easy for everyone to find and access our publications, 
thus reducing the time and money spent on access acquisition.
For \emph{findability}, 
we should consider collaborating on a \emph{living survey} of computational legal studies, 
i.e., a continuously updated community resource to help understand and navigate the CLS literature.
For \emph{accessibility}, 
let us embrace open access for papers, 
mirroring the principles of open data and open code (see Section~\ref{sec:sharing}).
While certainly not everything is great about open access publishing models \cite{ross2022dynamics,james2017free,mann2009open},  
and the common practice of having authors pay to publish the research they are paid to produce is far from ideal,
it is doubtlessly preferable to paywalled or print-only publications.
A simple start is to put final drafts on preprint servers such as \hyperref{https://arxiv.org/}{}{}{arXiv} or \hyperref{https://ssrn.com/}{}{}{SSRN}, 
but we should work towards making open access the rule, rather than the exception, also for published papers.

A culture of constructive community criticism goes beyond thorough engagement with related work, though. 
While most of us probably prefer discovering new knowledge over validating existing one, 
we should make it a habit to critically check and computationally reproduce each other's work.
\emph{Replication studies} are a crucial tool to keep the scientific record clean and deter sloppy research \cite{kahneman2014new},
and it should be entirely acceptable to question someone's results without questioning their scientific integrity.
There is nothing wrong with honest mistakes, and since we all know that they do happen, 
we should not dramatize them if we find them.

Replication studies are equally useful for educational purposes, 
as taking the authors' steps oneself is a great way to improve one's own understanding.
Therefore, conceptualizing and performing replication studies can be an excellent learning experience in the class room \cite{sodden2012class}.
While students have the opportunity to learn good scientific practices and produce replication studies in a systematic manner, 
they might even generate new insights from 
existing datasets or analyses \cite{king2000improving}. 
To make replications successful and avoid misunderstandings or errors in replications, 
both authors and replicators should adhere to a clearly defined etiquette when conducting them \cite{kahneman2014new}.
Most importantly, replication studies should be regarded as a collaborative effort, 
ideally leading to a joint publication by both authors and replicators in case a correction is needed.
As such, they should become an integral part of our research culture.

While we establish this culture, 
an easy way to practice constructive post-publication criticism without much of the otherwise looming social friction is self-correction by the authors.
To keep it simple and avoid clogging other publication venues, 
this could take the shape of a blog for \emph{matters arising}, with matters raised by (or in collaboration with) the original authors.
Such a blog could showcase minor improvements to methods used in past publications, 
as well as full-scale replications or extensions of studies with updated data.
Referencing this outlet in our papers could guide future readers towards updated insights.
By regularly revisiting our prior publications in a semi-formal setting, 
we might establish the right routine to avoid mistakes in the future.
Should this lead us to discover major flaws in our past work,
however, a formal correction in the original publication outlet is in order. 
Just as replication studies should be a precious part of our research routine, 
so should be corrections. 
Therefore: 
\begin{principle}
	Criticize collaboratively.
\end{principle}

%% file: text/outlook.tex
\section{Outlook}

In the previous sections, we have proposed six fundamental principles for constructive criticism in computational legal studies 
(deliberately skirting details that deserve separate in-depth debates, such as how to create informative figures), 
and we have sketched how these principles can be implemented.
When trying to anchor these principles in the research routines of the CLS community, however, 
we face the challenge that currently,
no publication outlet can provide a home for CLS as a whole.
Classical law reviews are not an option for the primary publication of computational legal studies 
because they overwhelmingly 
operate with a different understanding of peer review, 
have difficulties typesetting math or dealing with display items 
(especially when it comes to handling colors, layouting tables, or rendering figures), 
and cater to audiences less interested in methodological details.
However, they can still serve as secondary publication outlets to communicate narratives based on technical CLS publications that have undergone peer review elsewhere.

Publication outlets for empirical legal research 
or journals and conferences from legal informatics or core computer science are no alternatives:
On the one hand, not all computational legal studies are empirical, 
and those that \emph{are} often differ from classical empirical studies in both content and methodology.
For example, legal data science might concentrate on analyzing the law itself as a complex system, 
rather than its relationship to economics or society, 
and it might construct more complicated data representations 
and develop novel algorithms to analyze these representations, 
rather than rely on tabular data and established statistical tools.
On the other hand, legal informatics traditionally focuses on the formal foundations of legal technology,
which makes it a potential place for studies working towards computable law, 
but not for studies doing legal data science. 
Finally, turning to computer science conferences alone, 
with their self-organization by abstract problems and approaches, 
would prevent the unification of CLS as a field,
and none of the options discussed so far are set up 
to honor the intricacies of our transdisciplinary endeavor in the intersection of law and computer science.

We conclude that establishing a flagship publication outlet for the computational legal studies community deserves serious consideration. 
This flagship should be a digital native 
(to be honest, PDFs are not \emph{always} the best way to distribute research), 
it should have open science in its DNA, 
and it should accept \emph{publications}, not papers. 
As such, it ought to be \emph{truly} open access (i.e., free without making the authors pay for it), 
and it would need to enforce the best practices of rigorously peer-reviewed, computationally reproducible research.
This outlet could also host both the proposed 
\emph{matters arising} encouraging community self-correction, 
and the suggested \emph{living survey} of computational legal studies.
We encapsulate this idea in our seventh and last principle:
\begin{principle}
	Come together. 
\end{principle}